\documentstyle[aps,prl,multicol,epsf]{revtex}
\def\frac#1#2{{\textstyle{#1\over#2}}} %puts a small fraction in a 
\def\bra#1{\langle #1 |}
\def\ket#1{| #1\rangle}

\def\R{\hbox{\rm I \kern-5pt R}}

\def\ajou#1&#2(#3){\ \sl#1\bf#2\rm(19#3)}

\title{\rightline{ \small DAMTP-1999-51 \normalsize} \rightline{
\small quant-ph/9910087 \normalsize}
\rightline{} \centerline{
Unconditionally Secure Commitment of a Certified Classical Bit is 
Impossible} }

\author{ Adrian Kent}
\address{
Department of Applied Mathematics and 
Theoretical Physics, University of Cambridge,\\ 
Silver Street, Cambridge CB3 9EW, U.K.\\
} 

\date{October 20, 1999}
\begin{document}
\maketitle
\begin{abstract}
In a secure bit commitment protocol involving only classical physics, A  
commits either a $0$ or a $1$ to B.  If quantum information is 
used in the protocol, A may be able to commit a state of the form 
$\alpha \ket{0} + \beta \ket{1}$.  If so, she can also commit
mixed states in which the committed bit is entangled with 
other quantum states under her control.  
We introduce here a quantum cryptographic
primitive, {\it bit commitment with a certificate of classicality} (BCCC), 
which differs from standard bit commitment in that it guarantees 
that the committed state has a fixed classical value.  
We show that no unconditionally secure BCCC protocol 
based on special relativity and quantum theory exists.  
We also propose complete definitions of security for 
quantum and relativistic bit commitment.
\end{abstract}
\vskip10pt

\begin{multicols}{2}

\section{introduction}\label{introduction}

The discovery of secure quantum key 
distribution\cite{wiesner} and other applications of
quantum information raises the
question: which cryptographic tasks can be 
guaranteed secure by physical principles?  
One task which has been extensively investigated is
bit commitment(BC), an important cryptographic primitive with
several applications.  In this note, we distinguish 
between standard bit commitment and a stronger task, 
bit commitment with a certificate of classicality.
We show that, while the former can be guaranteed secure by 
physical principles, the latter cannot.
We also propose complete definitions of various 
types of physical security for bit commitment and 
related tasks.

\section{Physical security}

In the most commonly considered cryptographic scenario, A and B each
occupy disjoint laboratories which are treated as effectively
pointlike.  Each trusts the integrity of their own
laboratory but nothing outside; in particular, neither trusts
the other to accurately declare their location.  Under these assumptions, 
special relativistic signalling constraints cannot be relied on for
security, and the parties are effectively restricted
to protocols involving sequentially exchanged messages, each
party waiting to receive one message before sending the next.  

Special relativistic signalling constraints can, though, play a 
valuable r\^ole in ensuring security in cryptographic scenarios 
in which each party controls laboratories in two separate 
locations\cite{kentrel}.  While these laboratories must be near to mutually 
agreed coordinates, this can be tested within a given protocol: no trust
between the parties is required.  

We will neglect the possibility of 
cryptographic protocols whose security relies on general relativity, 
on quantum field theory as opposed to quantum mechanics, 
on the details of the standard model, and so on.
Similarly, we neglect the possibility of physically based
attacks relying on properties of these theories.  No practical 
protocols or attacks of these
types have so far been suggested.  

So far as physical security is concerned, then, cryptographic tasks
can presently be classified according to whether they can 
be securely implemented by using classical information and 
relying on special relativity, 
by using quantum information and neglecting special relativity,
or by relying both on the properties of quantum information and
on special relativity, or whether they cannot be securely implemented at all. 
We can divide cryptographic protocols into the following
classes.  
A {\it classical} protocol relies on the
exchange of classical information, while a {\it quantum}
protocol allows quantum information exchange.  
In a {\it non-relativistic} protocol,
the signalling constraints imposed by special relativity are 
neglected, usually because they make no essential difference
to the protocol's security.
In a {\it relativistic} protocol, special
relativity is taken as the underlying theory, and the parties
are located so that relativistic signalling constraints
play a crucial role by ensuring that some communications between
them are generated independently.
  
Similarly, a {\it classical attack} on a protocol is a cheating attempt by
one party which involves diverging from the defined protocol
and which can be described 
by standard non-relativistic or special relativistic classical
physics --- i.e. without using quantum information.
A {\it quantum attack} involves 
the transmission and/or manipulation of quantum information. 

\section{Bit commitment: definitions and results}\label{bcdefs}

We now recall the definitions of secure bit commitment in
classical cryptology. 
In a classical BC protocol A and B exchange
classical data in such a way that B obtains an encoding of a bit 
from A.  A need not necessarily know
which bit she has encoded: she could build a random bit 
generator and encoder and not inspect its operations.  
But if she follows the protocol, 
either $0$ or $1$ is committed, even if she does not know which.
For the protocol to be {\it classically
perfectly secure} against B, it must guarantee to A
that B cannot gain any information about a committed bit
until A chooses to reveal it. 
For it to be {\it classically perfectly secure} against A, 
it must guarantee to B that a committed  
bit is genuinely fixed between commitment and revelation.
That is, there must not be a cheating method allowing
A any chance of revealing the opposite bit to that 
committed.  More precisely (since A might have sent nonsense), 
perfect security requires that 
unless A committed a bit $a$ via the protocol, 
her probability of later revealing $a$ should be zero. 

A {\it security parameter} in a BC protocol is 
a positive integer parameter, $N$, such that security can be
increased by increasing $N$.  
The protocol is {\it classically secure} modulo certain assumptions if, when
the assumptions hold, the probability of A successfully 
cheating by classical attacks and the information available to B 
about the bit before revelation can simultaneously be 
made arbitrarily small by taking $N$ sufficiently large
It is {\it unconditionally classically secure} if its security is 
guaranteed if classical physics (in the case of relativistic protocols,
special relativistic classical physics) is valid.

In a quantum BC protocol, it may be possible for
A to commit a bit in an improper mixed state $\alpha \ket{0} \bra{0} + 
\beta \ket{1} \bra{1}$, by entangling the committed state with 
another state kept under her control.
She can do this, for example, by building a quantum computer 
which is programmed to follow the protocol for either of the
orthogonal input states $\ket{0}$ or $\ket{1}$ and  
inputting a superposition.  If she does so after entangling
a second system $\ket{~}_A$ with the input bit, by preparing 
$\alpha \ket{0} \ket{0}_A + \beta \ket{1} \ket{1}_A$, she
can keep this second system under her control throughout the
commitment, carry out a measurement of 
the observable with eigenbasis $\ket{0}_A , \ket{1}_A$ just
before revelation, and then reveal either a
$0$ or a $1$ to B depending on the measurement result. 
This possibility, whose security implications were first
pointed out and investigated by Brassard et al.\cite{bcms}, is not 
considered a security failure of a quantum 
BC protocol {\it per se}, although it can 
open up new cheating possibilities for A if the protocol
is part of a larger cryptographic scheme.  

So, we need revised definitions of security for quantum bit
commitment.  
Complete definitions of security for non-relativistic 
quantum BC does not seem to have
been set out yet, no doubt partly
because it is known\cite{mayersprl,mayerstrouble,mayersone,lochauprl,lochau}
that, under any
reasonable definition, unconditional security is unattainable
for non-relativistic quantum BC protocols.
However, precise definitions are needed to discuss security 
based on computational bounds for quantum BC 
and to discuss the security of relativistic quantum BC
protocols.  We propose definitions of 
security for non-relativistic quantum BC 
below, and extend them to the relativistic case. 

For a non-relativistic quantum BC protocol to be {\it
perfectly secure} against B, it must guarantee to A
that B can obtain no information about a committed bit
until A chooses to reveal it. 
For it to be {\it perfectly secure} against A, 
it must guarantee to B that A cannot act 
between commitment and revelation so as ever to
obtain some chance of choosing between revealing $0$ and $1$. 
More precisely, define $p_0 (t) $ be the probability of A
revealing to B a $0$ without giving him evidence
that she has violated the protocol, assuming that from time
$t$ onward she follows a strategy that maximizes her chances of 
doing so.  Define $p_1 (t)$ similarly.  Let $p(t) = p_0 (t) + p_1
(t)$.  Assume that B knows the commitment phase has
ended at $t=0$.
Then B must be guaranteed that,
however A acts after $t=0$, it must always be the case 
that $p(t) \leq 1$ for all $t>0$.  

A {\it security parameter} for a quantum BC protocol is 
a variable positive integer parameter, $N$, such that security can be
increased by increasing $N$.  
More precisely, we say a non-relativistic 
quantum BC protocol is {\it secure} modulo certain assumptions if, when
the relevant assumptions hold:

$(i)$ A is guaranteed that the information available to B during 
the protocol about the bit to be revealed can be bounded by $\epsilon$.

$(ii)$ B is guaranteed that, for every possible strategy of A's, 
the a priori probability of her making $p(t)>1$, for any $t>0$,
is uniformly bounded, whenever B is persuaded at $t=0$ that A has 
followed the commitment phase of the protocol. 
I.e., for any $\epsilon' > 0$, there is an $\epsilon''$ such 
that, regardless of the A's strategy, 
$${\rm Prob} \left( {\rm sup}_{t \geq 0} ( p(t) ) > 1 +
\epsilon'\right) < \epsilon'' \, .$$  

$(iii)$ For any $\epsilon' > 0$, 
$\epsilon$ and $\epsilon''$ can simultaneously be 
made arbitrarily small by increasing the security
parameter. 

A non-relativistic quantum BC protocol 
is {\it unconditionally secure} if its security is guaranteed
if non-relativistic quantum mechanics is valid. 

For relativistic quantum BC
protocols, we use similar definitions.
There must be some spacetime point $P$ at which
B is persuaded that A is committed.
For any space-time point $Q$, let $PC(Q)$ be the past light-cone
of $Q$ and $\bar{PC} (Q)$ be the rest of spacetime. 
We now define $p_0 (Q) $ to be the probability of A
revealing to B a $0$ without giving him any evidence
that she has violated the protocol, assuming that in
$\bar{PC}(Q)$ she follows a strategy 
that maximizes her chances of doing so, and $p_1 (Q)$ similarly, 
and set $p(Q) = p_0 (Q) + p_1 (Q)$.  
For perfect security we then require that, if B is persuaded of
a commitment at $P$, then however A acts 
in $\bar{PC}(P)$, we must have $p(Q) \leq 1$ for all $Q$ in $\bar{PC}(P)$. 
The definitions of security modulo assumptions 
are modified similarly. 
A relativistic quantum BC 
protocol is {\it unconditionally secure} if its security is 
guaranteed if quantum mechanics and special relativity are valid.

All non-relativistic classical BC schemes 
are in principle insecure, though very good practical security 
can presently be attained.  
Several quantum BC schemes have been 
proposed\cite{BBeightyfour,BCJL,brassardcrepeau}.
Again, for practical purposes, these schemes generally offer
very good security at present.  
However, in principle they are insecure. 
More generally, it was shown independently 
by Lo and Chau\cite{lochauprl,lochau}
and by Mayers\cite{mayersprl,mayerstrouble,mayersone} 
that no non-relativistic quantum BC schemes 
can be perfectly secure against both Bob and Alice. 
The restriction to non-relativistic schemes, though not made clear
in the cited papers, is crucial.  

The Lo-Chau-Mayers result 
was extended by Mayers\cite{mayersprl,mayerstrouble,mayersone} 
to give a proof of the 
general impossibility of unconditionally secure quantum
bit commitment.  Again, it should be noted that, despite initial
suggestions to the contrary\cite{mayersprl}, Mayers' proof 
applies only to non-relativistic schemes\cite{kentrel,kentrelfinite}. 
We refer here 
to the result that unconditionally secure non-relativistic quantum 
bit commitment is impossible as the NRQBC no-go theorem.  

More recently, several relativistic classical BC 
protocols have been proposed\cite{kentrel,kentrelfinite}.
These schemes are evidently secure against classical attacks 
and are all conjectured to be secure against quantum attacks.  
Though the first protocol proposed\cite{kentrel} requires an exponentially
increasing communication rate for its implementation, 
the later protocols\cite{kentrelfinite} can be implemented indefinitely
over communication channels of fixed capacity.  
No sharp optimality results are known; further refinements can
undoubtedly be made.
Our aim here, though, is not to examine the situation regarding
BC protocols in more detail, but to consider BCCC protocols.

\section{Bit commitment with a certificate of classicality} 

For many purposes, it would be desirable to have a BC
protocol which is guaranteed to behave like a classical protocol,
preventing A from exploiting the dangerous possibility\cite{bcms}
of committing a bit state which remains entangled.  
Formally, we define {\it bit commitment with a certificate
of classicality} (BCCC) to be a bit commitment in which the 
revelation of a bit $a$ guarantees that this particular bit was 
originally committed by A.
This does not necessarily imply that A was aware
of the value of $a$.  As with ordinary BC, she 
could arrange to remain ignorant, for instance by 
using a classical randomising device to prepare a 
proper mixed state and not inspecting the device.  

We say a BCCC protocol 
is {\it perfectly secure} against B if it guarantees to A
that B can obtain no information about the committed bit
until A chooses to reveal it. 
We say it is {\it perfectly secure} against A, 
if a revelation by A guarantees to B that the 
revealed bit was previously committed:
i.e., by some point in the protocol, a valid
commitment by A corresponds to her having 
input one of the states $\ket{0}, \ket{1}$ into some device
which generates her transmissions to B during the remainder
of the protocol, and a valid revelation of $a$ guarantees to B
that (precisely) the state $\ket{a}$ was input at that point.
As in the case of BC protocols, it must be possible to 
continue the commitment for
arbitrarily long between this point and the moment of revelation.  

A security parameter $N$ for a BCCC protocol 
is defined essentially as for a BC protocol.
Thus, we say a BCCC protocol is {\it secure} modulo 
certain assumptions if, when the assumptions hold, a revelation of $a$
guarantees that, with probability $1 - \epsilon$,
the fidelity of A's input state to the state $\ket{a}$ differed 
from $1$ by no more than $\epsilon'$, while
the information available to B about the bit 
during the protocol is no more than $\epsilon''$, 
where $\epsilon, \epsilon' , \epsilon''$ 
can simultaneously be made arbitrarily small by taking $N$ sufficiently large.
A non-relativistic BCCC protocol is {\it unconditionally secure} if 
its security is guaranteed if quantum mechanics is valid. 
A relativistic BCCC protocol is {\it unconditionally secure} if its 
security is guaranteed if quantum mechanics and special relativity are valid.

\section{Unconditionally secure BCCC is impossible} 

The main point of this paper is to show that no  
unconditionally secure BCCC protocol, relativistic or 
otherwise, exists.  We first give the proof,
and then comment.  

It is enough to show that no unconditionally secure 
relativistic protocol exists.
We prove this by contradiction.
Suppose that some unconditionally secure 
BCCC protocol existed.  Such a protocol might require 
A and B to occupy many sites, say $A_1 , .... , A_m$
and $B_1 , .... , B_n$.  Their locations may 
be time-dependent, provided that the relevant worldlines are 
timelike and that A and B's sites are always disjoint.    
We add a further constant velocity site $B_0$ for B, 
and use its stationary frame to define the time 
coordinate.  
Now suppose that A and B agree a large number $M$,
a much larger number $N_0 >> M$, and a large value 
$N_1$ for the security parameter.  They also fix
times of transmissions between their sites so as 
to run simultaneously $2 N_0$ BCCC 
protocols.  A chosen randomly and independently
$2 N_0$ bits, and commits those bits to B in the 
BCCC protocols.  

Regardless of the relative separations of the sites, B
can establish some time $t_c$ after the start of the 
protocols at which $B_0$ knows that, if A has followed
the BCCC protocols correctly, she is now committed (to the
extent that the security parameter prescribes).  
A is then required to send $B_0$, after time $t_c$, a sequence
of $N_0$ spin $1/2$ particles in one of the four spin states
$\ket{\uparrow}, \ket{\downarrow}, \ket{\leftarrow}, 
\ket{\rightarrow}$.  (The first two are eigenstates
of $\sigma_z$; the last two of $\sigma_x$.)  
This sequence is supposed to be correlated with the 
sequence of $N_0$ pairs of BCCC protocols given by the
first two, the second two, and so on.  In each case, if
the committed bits are respectively $(0,0)$, $(0,1)$, $(1,0)$, $(1,1)$, 
the state sent is supposed to be $\ket{\uparrow},
\ket{\downarrow}, \ket{\leftarrow}, 
\ket{\rightarrow}$.  

Once $B_0$ has received and stored the $N_0$ states, he randomly
picks $(N_0 -M )$ of them, and sends a message to A asking
her to reveal the corresponding commitments.  Some time $t_r > t_c$
is fixed such that the revealed bits are communicated back to $B_0$ from  
the $B_i$ by time $t_r$.  $B_0$ then checks that the revealed
bits do indeed characterise the spin-$1/2$ particle states, by
carrying out measurements in the appropriate basis for each of
the $(N_0 - M)$ particles.  If the tested particles pass these
checks, B accepts that the remaining $M$ particles are also 
(to very good approximation) in pure, unentangled 
eigenstates of $\sigma_x$ or $\sigma_z$.  
The corresponding $2 M$ BCCC protocols play no further
r\^ole, and are now suspended, without A
revealing the corresponding bits to B. 

A can now commit a single bit $a$ to B
via the following BC protocol.  Each of the 
$M$ untested spin-$1/2$ particles is (to good approximation)
in a $\sigma_z$ or $\sigma_x$ eigenstate.  Each
of these states is known to A but not to B. 
We let the variable $b$ be $x$ or $z$, and $\bar{b}$ the alternative.  
For a particle in a $\sigma_b$ eigenstate, A declares that, 
if the committed bit is $a$, the particle is a $\sigma_b$ eigenstate,
while if the committed bit is $\bar{a}$, the particle is a 
$\sigma_{\bar{b}}$ eigenstate. 

As the committed bits in the BCCC protocols were random,
these declarations give B no information about the bit
$a$.  But, since the BCCC protocols ensured that A would
almost certainly have been detected cheating unless
she sent the particles in (nearly) pure $\sigma_x$ or $\sigma_z$ 
eigenstates, this BC protocol does indeed commit her to $a$.  If she
follows the protocol, she can reveal $a$ 
by giving B the list of spin states, 
which he can check by measurements in the appropriate bases.  
But if she is dishonest, for one of the two possible
commitments, say $a_f$, at least $M/2$ of declarations are false.  
Her probability of persuading B that the committed bit was $a_f$,
by producing for him a list of spin states which pass his tests,
is approximately $(1/2)^{M/2}$.  I.e., for sufficiently
large values of $M$ and the other parameters, B will almost 
certainly detect a cheating attempt. 

Now, if the BCCC protocols were unconditionally secure, then the ensuing
BC protocol is also unconditionally secure.  In other words, by
combining these protocols into one, we have an unconditionally
secure relativistic BC protocol {\it with the property that 
after a finite number of transmissions the commitment
is complete}.  While unconditionally secure relativistic BC
protocols exist\cite{kentrel,kentrelfinite}, these protocols
require that transmissions continue indefinitely up
to revelation.  The same argument  
used to establish the NRQBC no-go 
theorem\cite{mayerstrouble,mayersone,lochauprl,lochau} shows 
that no finite unconditionally secure relativistic BC protocol 
exists.  Hence unconditionally secure BCCC is impossible.  

\section{Discussion} 

This result re-emphasizes that classical
cryptographic relations cannot naively be transferred into
the quantum realm.  In classical cryptology, non-relativistic
or relativistic, there is no distinction between BC and BCCC: 
in quantum relativistic cryptology, BC can be implemented with
unconditional security, while BCCC cannot.  

A less direct argument for the impossibility of 
unconditionally secure BCCC follows from results of Yao\cite{yao}, 
which imply that unconditionally secure oblivious 
transfer (OT) could be built from unconditionally secure BCCC. 
Since non-relativistic BC can straightforwardly be constructed from OT, 
we again reach a contradiction with the NRQBC no-go theorem.  

Note, finally, that while unconditionally secure BCCC is impossible,
BCCC schemes with security based on computational
assumptions are certainly possible.  Most standard classical BC 
schemes that are perfectly secure against A --- for example, those
based on factorisation or obtaining a discrete logarithm --- have
this property.  
An interesting recent quantum BC proposal by Salvail\cite{salvail} also
has this property.  It would be very interesting to understand whether 
BCCC schemes can be built with security based on assumptions which can
confidently be relied upon in a future quantum technological era.  

\vskip5pt
\leftline{\bf Acknowledgments}

I thank the Royal Society for financial support and the Oxford Centre
for Quantum Computation for much appreciated hospitality.  
I am very grateful to Gilles Brassard, Claude Cr\'epeau, Dominic
Mayers, Louis Salvail for helpful discussions.

\end{multicols}


\begin{thebibliography}{99}

\bibitem{wiesner}
S.~Wiesner, SIGACT News {\bf 15} (1983) 78.
\bibitem{BBeightyfour}
C. H. Bennett and G. Brassard, in {\it Proceedings of IEEE
 International Conference on Computers, Systems and Signal Processing} (IEEE,
 New York, 1984), p.~175.
\bibitem{BCJL}
G.~Brassard, C.~Cr\'{e}peau, R.~Jozsa and D.~Langlois, in
 {\it Proceedings of the 34th Annual IEEE Symposium on the Foundation of
 Computer Science} (IEEE Comp. Soc., Los Alamitos, California, 1993), p.~362.
\bibitem{brassardcrepeau}
G.~Brassard and C.~Cr\'{e}peau, 
in {\it Advances in Cryptology:
 Proceedings of Crypto'90}, Lecture Notes in Computer Science Vol 537
(Springer-Verlag, Berlin, 1991), p.~49.
\bibitem{lochauprl}
H.-K.~Lo and H.~Chau,  Phys. Rev. Lett. { \bf 78}  (1997)
3410.
\bibitem{mayersprl}
D.~Mayers,  Phys. Rev. Lett. {\bf 78} (1997) 3414.
\bibitem{mayerstrouble}
D.~Mayers, quant-ph/9603015.
\bibitem{lochau}
H.-K.~Lo and H.~Chau,  Physica D {\bf 120} (1998) 177.
\bibitem{lo}
H.-K.~Lo,  Phys. Rev. A {\bf 56} (1997) 1154.
\bibitem{mayersone}
D.~Mayers, in {\it Proceedings of the Fourth Workshop on
 Physics and Computation} (New England Complex System Inst., Boston, 1996),
 p.~226.
\bibitem{bcms}
G.~Brassard, C.~Cr\'epeau, D.~Mayers and L.~Salvail, 
quant-ph/9806031.
\bibitem{kentrel}
A.~Kent, Phys. Rev. Lett. 83 (1999) 1447-1450.
\bibitem{kentrelfinite} 
A.~Kent, Secure Classical Bit Commitment over Finite Channels, 
quant-ph/9906103, submitted to J. Cryptology. 
\bibitem{yao} 
A.~Yao, in {\it Proceedings to the 26th Symposium on the Theory
of Computing}, June 1995, pp. 67-75. 
\bibitem{salvail} 
L.~Salvail, in {\it Proceedings of Crypto'98}, 
Lecture Notes in Computer Science Vol 1462 
(Springer-Verlag, Santa-Barbara, 1998) pp.~338-353.


\end{thebibliography}
\end{document}